\documentclass[aps ,12pt,tightenlines
,twoside]{revtex4}

\usepackage{amsmath}
\usepackage{amstext}
\usepackage{amssymb}
\usepackage[english]{babel}
\usepackage[latin2]{inputenc}
\begin{document}

\author{Andrzej Grudka}
\email{agie@amu.edu.pl}

 \author{Ravindra W. Chhajlany}
 \email{chhajl@hoth.amu.edu.pl}

\affiliation{Faculty of Physics, Adam
  Mickiewicz University, Umultowska 85, 61-614 Pozna\'n, Poland}
\date{\today} \title{Two-way teleportation}

\begin{abstract}
  A protocol for teleporting two qudits simultaneously in opposite
  directions using a single pair of maximally entangled qudits is
  presented. This procedure works provided that the product of
  dimensions $d_{1}$ and $d_{2}$ of the two qudits to be teleported
  does not exceed the dimension $d$ of the individual qudits in the maximally
  entangled pair.
\end{abstract}

\maketitle

Quantum teleportation, the process of transferring complete
information about a quantum state from a sender to a receiver, is one
of the fundamental features of quantum information. In the original
setting \cite{bennett+93}, Alice and Bob utilize a shared maximally
entangled pair of qubits to teleport a qubit in some unknown state
from Alice to Bob.  This is implemented by local operations by Alice
and Bob supplemented by classical communication between them. This
scheme is easily generalized to the case of qudits ($d$-dimensional
quantum systems) \cite{bennett+93, stenholm+98}.  In order to teleport
a $d$-dimensional qudit, Alice and Bob must share a pair of at least
$d$-dimensional qudits in the maximally entangled state. Quantum
teleportation between multiple parties using qubits \cite{karlsson+98,
  hillery+99} and qudits \cite{murao+00, ghiu03} has also been widely
studied. In all teleportation protocols, the pair of maximally
entangled qudits plays the role of a uni-directional quantum channel.

In this article we analyze two-way quantum teleportation, {\it i.e.}
simultaneous teleportation of two qudits in two opposite directions,
as introduced in \cite{grudka03}. Suppose that Alice possesses a
$d_{1}$-dimensional qudit in the state $|\alpha \rangle_{1}=
\sum_{i=0}^{d_{1} -1} \alpha_{i} |i \rangle_{1}$, while Bob has a
$d_{2}$-dimensional qudit in the state $|\beta \rangle_{2}=
\sum_{i=0}^{d_{2} -1} \beta_{i} |i \rangle_{2}$. The qudits $1$ and
$2$ will henceforth be called teleportee qudits.  Additionally Alice
and Bob share a maximally entangled pair of $d$-dimensional qudits,
which serves the purpose of a quantum channel, in the state
\begin{gather}
|\Psi  \rangle = \frac{1}{ \sqrt{d}} \sum_{i=0}^{d-1} |i \rangle_{c1}|i
 \rangle_{c2} ,
\label{Rxwa}
\end{gather}
such that $d_{1}d_{2} \leq d$. In accordance with the notation, the
qudits in the state $|\Psi \rangle $ will be referred to as channel
qudits with the first qudit belonging to Alice and the second to Bob.
By applying the standard teleportation protocol of Bennett {\it et
  al.}, Alice and Bob can manage to teleport only one of the two
states $|\alpha \rangle $ or $|\beta \rangle $ using the channel
described by Eq.(\ref{Rxwa}). We construct an explicit scheme enabling
the teleportation of $|\alpha \rangle_{1}$ from Alice to Bob and
$|\beta \rangle_{2}$ from Bob to Alice using the shared maximally
entangled pair in the state $|\Psi \rangle $.


Our teleportation scheme consists of three main stages: 1) tailoring of
the quantum channel, 2) encoding of information about the states to be
teleported into the channel qudits and 3) decoding information in the
channel into qudits in product form.  We divide our considerations
into two cases, {\it viz}. $d_{1} d_{2} = d$ and $d_{1} d_{2} <d$. 

We first consider the former case, where the quantum channel is
characterized by the dimension $d = d_{1} d_{2}$. In this case, the
channel is already in the desired state. The first step of the
encoding stage consists of Alice projecting her two qudits into one of
$d_{1}$ orthogonal $d$-dimensional subspaces of the original
$d_{1}d$-dimensional space, by measuring the projectors
\begin{gather}
  P_{k_{1}} = \sum_{j_{1}=0}^{d_{1} -1}\sum_{j_{2}=0}^{d_{2} -1}
  |j_{1}\rangle_{1}|(j_{1} \oplus_{1} k_{1} ) + j_{2}d_{1}
  \rangle_{c1} \; _{c1}\langle (j_{1} \oplus_{1} k_{1}) + j_{2} d_{1}|
  _{1}\langle j_{1}|, \\ k_{1} = 0, \ldots, d_{1}-1,\nonumber
\label{Rpwa}
\end{gather}
where $\oplus_{1}$ denotes addition modulo $d_{1}$. Bob performs an
analogous measurement on his qudits with $d_{2}$ outcomes, $k_{2} =
0, \ldots, d_{2}-1$,  described by
\begin{gather}
  P_{k_{2}} = \sum_{j_{1}=0}^{d_{1} -1}\sum_{j_{2}=0}^{d_{2} -1}
  |j_{2}\rangle_{2}|j_{1} + (j_{2} \oplus_{2} k_{2}) d_{1}
  \rangle_{c2} \;_{c2}\langle j_{1} + (j_{2} \oplus_{2} k_{2}) d_{1}|
  _{2}\langle j_{2}|,
\label{Rqwa}
\end{gather}
with $\oplus_{2}$ denoting addition modulo $d_{2}$. If Alice and Bob
obtain the results $k_{1}$ and $k_{2}$ respectively, then the joint
state of all four qudits considered is
\begin{gather}
  |\psi_{1}\rangle = \sum_{j_{1}=0}^{d_{1}-1} \sum_{j_{2}=0}^{d_{2}-1}
  \alpha_{j_{1}} \beta_{j_{2}} |j_{1}\rangle_{1} |(j_{1} \oplus_{1}
  k_{1} ) + (j_{2} \oplus_{2} k_{2})d_{1}\rangle_{c1}
  |j_{2}\rangle_{2} |(j_{1} \oplus_{1} k_{1} ) + (j_{2} \oplus_{2}
  k_{2})d_{1}\rangle_{c2}.
\label{Rswa}
\end{gather}
Next, Alice and Bob inform each other about the results of their local
measurements. Using these results, they perform the following
operation on their channel qudits
\begin{gather}
  |(j_{1} \oplus_{1} k_{1}) + (j_{2} \oplus_{2}
  k_{2})d_{1}\rangle_{cr} \rightarrow |j_{1} + j_{2 }
  d_{1}\rangle_{cr},
\label{Rzwa}
\end{gather}
where $r=1, 2$ refer to Alice's and Bob's qudits respectively.  The
state $|\psi_{1}\rangle $ is transformed to
\begin{gather}
  |\psi_{2}\rangle = \sum_{j_{1}=0}^{d_{1}-1}
  \sum_{j_{2}=0}^{d_{2}-1} \alpha_{j_{1}} \beta_{j_{2}} |j_{1}\rangle_{1}
  |(j_{1} + j_{2} d_{1}\rangle_{c1} |j_{2}\rangle_{2}|j_{1} + j_{2}
  d_{1}\rangle_{c2}.
\label{Rvwa}
\end{gather}
Now Alice and Bob perform the following unitary operation on their
respective pairs of qudits:
\begin{gather}
|j_{r}\rangle_{r} |j_{1} + j_{2} d_{1}\rangle_{cr} \rightarrow |0
 \rangle_{r} |j_{1} + j_{2} d_{1}\rangle_{cr},
\label{Rjxa}
\end{gather}
that leaves the first qudit of each pair in a standard state $|0
\rangle_{r}$ ($r=1,2$) and the channel qudit unchanged. 
In this way, entire information about the teleportee states is encoded
into the quantum channel, which is in the state
\begin{gather}
  |\psi_{4} \rangle = \sum_{j_{1}=0}^{d_{1}-1}\sum_{j_{2}=0}^{d_{2}-1}
  \alpha_{j_{1}} \beta_{j_{2}} |j_{1} + j_{2}d_{1}\rangle_{c1} |j_{1}+
  j_{2}d_{1}\rangle_{c2}.
\label{R4wa}
\end{gather}
while the other two qudits can be discarded.  The two parties now
share the information about both  states $|\alpha \rangle $ and
$|\beta \rangle $.

To complete the teleportation procedure, Alice and Bob have to decode
the information present in $|\psi_{4}\rangle $. To do this, Alice
performs a $d_{1}$-dimensional Fourier transform described below:
\begin{gather}
  |j_{1} + j_{2} d_{1}\rangle_{c1}\rightarrow \frac{1}{\sqrt{d_{1}}}
  \sum_{m_{1}=0}^{d_{1}-1} \omega_{d_{1}}^{m_{1}j_{1}} |m_{1} + j_{2}
  d_{1}\rangle_{c1},
\label{R5wa}
\end{gather}
where $\omega_{d_{1}} = \exp[i 2 \pi /d_{1}]$ is complex root of
unity.  Bob performs a similar Fourier transform on his channel qudit
\begin{gather}
  |j_{1} + j_{2} d_{1}\rangle_{c2}\rightarrow \frac{1}{\sqrt{d_{2}}}
  \sum_{m_{2}=0}^{d_{2}-1} \omega_{d_{2}}^{m_{2}j_{2}} |j_{1} + m_{2}
  d_{1}\rangle_{c2}. 
\label{R6wa}
\end{gather}
The state of the two channel qudits is transformed in this way into 
\begin{gather}
  |\psi_{5}\rangle = \frac{1}{\sqrt{d} } \sum_{m_{1}=0}^{d_{1}-1}
  \sum_{j_{1}=0}^{d_{1}-1} \sum_{m_{2}=0}^{d_{2}-1}
  \sum_{j_{2}=0}^{d_{2}-1} \alpha_{j_{1}} \beta_{j_{2}}
  \omega_{d_{1}}^{m_{1}j_{1}}\omega_{d_{2}}^{m_{2}j_{2}} |m_{1} +
  j_{2}d_{1}\rangle_{c1} |j_{1}+ m_{2}d_{1}\rangle_{c2}.
\label{R7wa}
\end{gather}
Alice and Bob disentangle their qudits by performing the projections
\begin{gather}
  Q_{m_{1}} = \sum_{j_{2}=0}^{d_{2}-1} |m_{1} + j_{2} d_{1}\rangle
  \langle m_{1} + j_{2} d_{1}|, \; m_{1}= 0, \ldots, d_{1}
\label{R8wa}
\end{gather}
and
\begin{gather}
  Q_{m_{2}} = \sum_{j_{1}=0}^{d_{1}-1} |j_{1} + m_{2} d_{1}\rangle
  \langle j_{1} + m_{2} d_{1}|, \; m_{2}= 0, \ldots, d_{2}.
\label{R9wa}
\end{gather}
on their respective qudits.  Any pair of results $m_{1}, m_{2}$ of
these local measurements occurs with equal probability $d^{-1}$
yielding the product state
\begin{gather}
|\psi_{6}\rangle = \sum_{j_{1}=0}^{d_{1}-1} \sum_{j_{2}=0}^{d_{2}-1}
\alpha_{j_{1}} \beta_{j_{2}} \omega_{d_{1}}^{m_{1}j_{1}}
\omega_{d_{2}}^{m_{2}j_{2}}
|m_{1} + j_{2} d_{1}\rangle_{c1} |j_{1} + m_{2}
 d_{1}\rangle_{c2} .
\label{R0wa}
\end{gather}
Alice and Bob communicate the results of the measurements $Q_{m_{1}}$
and $Q_{m_{2}}$ (Eqs.(\ref{R8wa}) and (\ref{R9wa})) to one another and
perform the unitary operations $U_{1}$ and $U_{2}$:
\begin{gather}
U_{1} |m_{1} + j_{2} d_{1}\rangle_{c1} = \omega_{d_{2}}^{-m_{2}
 j_{2}} |j_{2} d_{1}\rangle_{c1},
\label{Raxa}
\end{gather}
\begin{gather}
U_{2} |j_{1} + m_{2} d_{1}\rangle_{c2} = \omega_{d_{1}}^{-m_{1}
 j_{1}} |j_{1} \rangle_{c2}
\label{Rbxa}
\end{gather}
on their qudits.  The  channel qudits end up in the state
\begin{gather}
  |\psi_{7}\rangle = \sum_{j_{2}=0}^{d_{2}-1} \beta_{j_{2}} | j_{2}
  d_{1}\rangle_{c1} \sum_{j_{1}= 0}^{d_{1}-1} \beta_{j_{1}}
  |j_{1}\rangle_{c2} = (V_{1}^{\dagger} |\beta \rangle_{c1}) |\alpha
  \rangle_{c1}.
\label{Rcxa}
\end{gather}

Bob's qudit is now in the desired state $|\alpha \rangle $ while
Alice's qudit completely encodes the state $|\beta \rangle$. Alice
only has to perform a rotation of basis $V_{1}|j_{2} d_{1}\rangle_{c1}
= |j_{2}\rangle_{c1}$ to exactly retrieve the original state. In this
manner, teleportation in two opposite directions is achieved.

Notice that teleportation of both states in this process can
essentially be conducted simultaneously. This process is successful
with unit efficiency since it is implemented using only unitary
operations and complete measurements.

Now, consider the case when $d_{1}d_{2} = d' < d $. In this case, the
protocol presented above cannot be directly applied unless the channel
is a maximally entangled pair of $d'$-dimensional states.  This reason
behind this being that the projections applied do not span the entire
local spaces of the given qudits, which in turn may lead to a loss of
information due to the possible projection on to some complementary
space.  The conversion of the state of the given channel
(Eq.(\ref{Rxwa})) into the required form can be conducted beforehand
by Alice and Bob in  a ``tailoring'' stage, as discussed below.

Alice (or Bob) can prepare an ancillary $d'$-dimensional qudit in the
state 
\begin{gather}
|\phi  \rangle = \frac{1}{\sqrt{d'} } \sum_{i=0}^{d'-1} |i \rangle_{0},
\label{Rdxa}
\end{gather}
{\it e.g.} by applying the $d'$-dimensional Fourier transform to the
state $|0 \rangle_{0}$. The joint state of the ancilla and channel
qudits is $|\phi \rangle |\Psi \rangle $. Alice performs a measurement
having $d$ outcomes defined by the projection operators
\begin{gather}
  R_{k} = \sum_{n=0}^{d'-1}|n \rangle_{0}|n+k \rangle_{c1}
  \;_{c1}\langle n+k |   \;_{0}\langle n |,
 \label{Rexa} 
\end{gather}
where $k=0,1, \ldots d-1$ and all sums are taken modulo $d$. The state
$|\phi \rangle |\Psi \rangle $ is projected with equal probability
$d^{-1}$ into one of the states
\begin{gather}
  |\varphi_{k}\rangle =\frac{1}{\sqrt{d'} }\sum_{n=0}^{d'-1} |n
  \rangle_{0} |n+k \rangle_{c1}|n+k \rangle_{c2},
\label{Rfxa}
\end{gather}
Alice unitarily separates the ancilla from her channel qudits by
the operation
\begin{gather}
|n \rangle_{0} |n+k \rangle_{c1} \rightarrow |0 \rangle_{0} |n+k \rangle_{c1}.
\label{Rkxa}
\end{gather}
Alice sends the result $k$ to Bob, and both parties perform the
subtraction operation $|i \rangle_{cr} \rightarrow |i-k \rangle_{cr}$
leaving the channel qudits in the maximally entangled states
 \begin{gather}
   |\Psi ' \rangle = \frac{1}{ \sqrt{d'}} \sum_{i=0}^{d'-1} |i
   \rangle_{c1} |i \rangle_{c2}.
\label{Rixa}
\end{gather}
This pair of qudits can now be used as a channel for teleporting two
qudits in opposite directions, by performing steps 2) and 3) as
described earlier.

A general protocol utilizing a maximally entangled state as a
bi-directional channel for teleportation has been presented. The only
condition for successful teleportation is that the information to be
exchanged occupies a space of at most as high a dimension as that
occupied by the channel.

One of us (A. G.) would like to thank the State Committee for
Scientific Research for financial support under Grant No. 0 T00A 003
23.

\end{document}